\documentclass[aps,prb,twocolumn,groupedaddress]{revtex4-2}
\usepackage{graphicx}
\usepackage[utf8]{inputenc}
\usepackage{amsmath,amssymb}
\usepackage{booktabs}
\begin{document}
\title{Entropy per Domain-Wall Step and the Structure of Criticality
       in the $q$-State Potts Model}
\author{David Vaknin}
\affiliation{Department of Physics and Astronomy, Iowa State University,
             Ames, Iowa 50011, USA}
\date{\today}
\begin{abstract}
We develop a geometric framework for Potts criticality based on the
free energy per domain-wall step, $f_{\rm step} = E_{\rm step} -
Ts_{\rm step}$, where the bond energy $E_{\rm step}=(z-2)J$ and the
configurational entropy $s_{\rm step}\equiv\ln\lambda$ are both derived
independently from lattice geometry.  The critical temperature follows
from the balance $f_{\rm step}=0$, giving $T_c = E_{\rm step}/s_{\rm
step}$, without explicit evaluation of the partition function.
Two independent lattice properties govern the structure of the description.
\emph{Self-duality}: on the square lattice, which is its own dual, the
minimal two-state transfer matrix gives $\lambda=1+\sqrt{q}$ and the
free-energy balance independently reproduces the exact critical
temperature.  For non-self-dual lattices the matrix must be built on
the dual geometry where domain walls propagate; the spin-model $T_c$
then follows from the exact duality relation between the dual lattice
pair, illustrated for the triangular--honeycomb case.
\emph{Bipartiteness}: on bipartite lattices, topology and color entropy
can be separated at the level of the minimal transfer matrix; on
non-bipartite lattices a
Junction state --- a point where three or more domains meet --- is
required and couples them irreducibly at the level of the minimal
transfer matrix.
Motivated by the square-lattice result, we propose the geometric heuristic
$\lambda_{\rm ansatz}=m+q^p$, where $m$ encodes topological persistence
and $p$ encodes color entropy per step, with no fitted parameters.
Applied to the simple cubic lattice this gives $T_c$ within less than
$1\%$ of simulation for $q=2$.  The Junction state is identified as
the topological object responsible for the Wannier--Baxter
antiferromagnetic frustration entropy; bipartiteness suppresses it,
and the enlarged state space, irreducible characteristic polynomial,
and extensive frustration entropy all vanish simultaneously.
\end{abstract}
\maketitle
\section{Introduction}
\label{sec:intro}
The $q$-state Potts model~\cite{Wu1982} is a central model of
statistical mechanics. In two dimensions, Kramers--Wannier
duality~\cite{KW1941} locates the critical point of the square-lattice
model at the self-dual condition $v^2=q$, assuming a unique transition;
Onsager's exact solution~\cite{Onsager1944} of the $q=2$ case confirmed
this and provided the full free energy. Exact results for general $q$
followed from the star-triangle (Yang--Baxter) relation~\cite{Baxter1982}.
These approaches all rely on global transformations of the partition
function.
Here we adopt a complementary viewpoint based on the growth of
domain-wall configurations.  The central quantity is the
\emph{configurational entropy density per interface step},
\begin{equation}
    s_{\rm step} \equiv \ln\lambda
    = \lim_{N\to\infty}\frac{1}{N}\ln\Omega_N,
    \label{eq:sstep}
\end{equation}
where $\Omega_N$ is the number of allowed interface histories of length
$N$ and $\lambda$ is the corresponding exponential growth rate.  The
quantity $\lambda$ should be understood as the spectral radius of a
local transfer operator acting on a reduced interface state space.  It
is therefore a large-deviation rate for \emph{open} local interface
histories, not the full density of states of the spin model and not the
closed-loop ensemble entering the exact partition function.
The working hypothesis of the present paper is that the critical point
is controlled, to leading exponential accuracy, by the competition
between the energetic cost of extending an interface and the
combinatorial growth rate of such interface histories.  A wall of $N$
steps carries energy $U=N\,E_{\rm step}$, where $E_{\rm step}=(z-2)J$
is the minimal energy cost per forward propagation of an interface,
corresponding to the number of broken bonds per step on a lattice of
coordination $z$ (the factor $z-2$ excludes the bond pointing backward
and the bond pointing along the wall).  The entropy is
$S=N\,s_{\rm step}$, giving the coarse-grained free energy per step
\begin{equation}
    f_{\rm step} = E_{\rm step} - T\,s_{\rm step}.
\end{equation}
Setting $f_{\rm step}=0$ yields the critical condition:
\begin{equation}
    T_c = \frac{E_{\rm step}}{\ln\lambda}.
    \label{eq:Tc}
\end{equation}
This is a Peierls-type criterion~\cite{Peierls1936} formulated in terms
of a reduced domain-wall ensemble.  It is not a derivation of the full
partition function, and it neglects loop-closure constraints, wall--wall
interactions, and subleading correlations.  Its nontrivial content is
that, in favorable cases, the leading exponential rate $\lambda$ is
sufficient to recover the correct critical structure: closure constraints
contribute only sub-exponential corrections and therefore do not affect
$\lambda$.  The estimate is exact on the square lattice, where
self-duality provides an independent algebraic fix on $T_c$; on other
lattices it is a controlled Markovian approximation.
Two independent lattice properties organize the results.
\emph{Self-duality} determines whether the domain-wall counting can be
identified directly with the spin-model critical condition.  In two
dimensions, domain walls are edges of the dual lattice by construction.
For the square lattice this distinction collapses because the lattice is
self-dual, and the counting directly reproduces the exact monomial
critical condition.  For non-self-dual lattices, however, the transfer
matrix must be built on the dual-lattice geometry where domain walls
actually propagate; the spin-model critical point is then obtained only
after an additional exact duality step, illustrated for the
triangular--honeycomb pair in Sec.~\ref{sec:hc}.
\emph{Bipartiteness} controls the structure of the reduced interface
state space.  On bipartite lattices, topology and color multiplicity
can be separated at the level of the minimal transfer matrix.  On
non-bipartite lattices, this separation fails: a Junction state,
representing the meeting of three or more domains, becomes unavoidable
and couples geometry to color entropy irreducibly, as detailed in
Secs.~\ref{sec:triangular} and~\ref{sec:junction}.
The goal of the paper is therefore not to replace exact solutions, but
to isolate the minimal geometric ingredients that control the leading
entropy of interfaces.  In this language, self-duality explains when
the reduced counting can become exact, while bipartiteness explains when
topology and color entropy remain separable.
\section{Effective Free-Energy Balance and the Geometric Ansatz}
The energy cost per interface step is
\begin{equation}
    E_{\rm step} = (z-2)J,
    \label{eq:Estep}
\end{equation}
as introduced in Sec.~\ref{sec:intro} above.  Throughout this paper the bond fugacity is
\begin{equation}
    v = e^{2J/T} - 1,
    \label{eq:fugacity}
\end{equation}
consistent with the temperature normalization adopted here.  Some
references, including Wu~\cite{Wu1982} and Baxter~\cite{Baxter1982}, use
$v=e^{K}-1$ with $K=\beta J$; our convention corresponds to $K=2J/T$.
All critical temperatures have been verified against standard tabulated
values.
The minimal transfer matrices constructed below provide a definite
estimate of $s_{\rm step}$ within the Markovian approximation of the
local state space.
Motivated by the square-lattice result and the geometric argument for
the simple cubic developed in Appendix~\ref{app:sc}, we hypothesize
that for bipartite orthogonal lattices the entropy per step takes the
form
\begin{equation}
    s_{\rm step} = \ln(m + q^p),
    \qquad\text{equivalently,}\qquad
    \lambda = m + q^p,
    \label{eq:ansatz}
\end{equation}
where $m$ is a purely topological integer independent of $q$, encoding
the number of ways an interface can continue without introducing a new
color degree of freedom, and $p$ encodes how color entropy enters per
step.  We argue below that bipartiteness is required to keep $m$
independent of $q$, and orthogonality to keep the domain-wall counting
simple enough for the form to hold.  This hypothesis is exact on the
square lattice, where self-duality independently fixes $T_c$, and
serves as a parameter-free approximation for the simple cubic, where
no exact result is available.
\section{Square Lattice: Exact Result and Self-Duality}
\subsection{Counting domain-wall configurations}
On the square lattice ($z=4$) a domain wall is a path on the dual
lattice that separates regions of different Potts color.  At each
step the wall is in one of two states: \textbf{Bulk} ($a_n$), where
no wall is active and the interface has not yet opened, and
\textbf{Wall} ($b_n$), where an active wall separates two distinct
Potts colors.
The allowed transitions per step are:
\begin{itemize}
  \item Bulk $\to$ Bulk: the interface continues within a single
    domain ($1$ way).
  \item Bulk $\to$ Wall: a new wall opens; the new domain chooses
    one of $q$ available Potts colors, giving $q$ ways.
  \item Wall $\to$ Bulk: the wall closes and the color is absorbed
    back into the surrounding uniform domain ($1$ way).
  \item Wall $\to$ Wall: the wall continues ($1$ way).
\end{itemize}
These weights are all integers and encode only local geometry ---
no Boltzmann factors, no thermodynamic assumptions.  Collecting them
into a transfer matrix:
\begin{equation}
    M = \begin{pmatrix} 1 & 1 \\ q & 1 \end{pmatrix},
    \label{eq:Msq}
\end{equation}
where rows index the destination (next) state and columns index the
source (current) state, so that $M$ acts by left-multiplication on
the column vector $(a_n,b_n)^T$.
The dominant eigenvalue of $M$ is $\lambda = 1+\sqrt{q}$, giving
\begin{equation}
    s_{\rm step} = \ln(1+\sqrt{q}),
    \qquad
    T_c = \frac{2J}{\ln(1+\sqrt{q})}.
    \label{eq:sq_result}
\end{equation}
This corresponds to the hypothesis~\eqref{eq:ansatz} with $m=1$ and $p=1/2$.
The result is consistent with the exact critical temperature guaranteed
by Kramers--Wannier duality~\cite{Wu1982}; the cases $q\leq4$ in
Table~\ref{tab:sq_Tc} lie in the continuous-transition regime.  At
$q=1$, $\lambda=2$ gives $p_c=1/2$, the exact bond-percolation
threshold~\cite{Sykes1964}.
\begin{table}[h]
\caption{Critical temperatures for the square lattice ($z=4$).
All values for $q\leq4$ are exact via Kramers--Wannier duality.}
\label{tab:sq_Tc}
\begin{ruledtabular}
\begin{tabular}{cccc}
$q$ & $\lambda=1+\sqrt{q}$ & $s_{\rm step}=\ln\lambda$ & $T_c/J$ \\
\hline
1 & 2.000 & 0.6931 & 2.8854 \\
2 & 2.414 & 0.8814 & 2.2692 \\
3 & 2.732 & 1.0051 & 1.9898 \\
4 & 3.000 & 1.0986 & 1.8205 \\
\end{tabular}
\end{ruledtabular}
\end{table}
\subsection{Why self-duality makes the counting exact}
The Kramers--Wannier critical condition is $v^2=q$, i.e., $v=\sqrt{q}$
with $v=e^{2J/T}-1$ --- a \emph{pure monomial}.  The counting matrix
eigenvalue gives $v=\lambda-1=\sqrt{q}$, which satisfies this condition
exactly.
This works because the square lattice is self-dual: domain walls live on
the dual lattice, which is again a square lattice.  The counting matrix
therefore describes a process on the same geometric object as the spin
model, so $s_{\rm step}$ directly encodes the critical fugacity.  For
non-self-dual lattices the dual is a different lattice, and connecting
$s_{\rm step}$ to $T_c$ requires the explicit duality relation
$v\cdot v^*=q$.
\section{Triangular Lattice: Junction State}
\label{sec:triangular}
The triangular lattice ($z=6$) is non-bipartite: three distinct Potts
colors can meet simultaneously at a single vertex.  Domain walls on the
triangular lattice are paths on the dual honeycomb lattice, whose
vertices are trivalent: one edge arrives and two edges leave at each
step.  A minimal description therefore requires a third state beyond
Bulk and Wall, namely the \textbf{Junction} ($j_n$), representing a
Y-shaped branching point where three domain walls meet and three
distinct Potts colors are simultaneously present.
The allowed transitions per step, enumerated on the trivalent dual
vertex, are:
\begin{itemize}
  \item Bulk $\to$ Bulk: the step stays in a single domain ($1$ way).
  \item Bulk $\to$ Wall: a wall opens along one forward edge; the
    color pair is absorbed into the Wall state ($1$ way).
  \item Bulk $\to$ Junction: impossible without an active wall already
    present ($0$ ways).
  \item Wall $\to$ Bulk: the wall closes; the region ahead takes any
    of $q$ colors ($q$ ways).
  \item Wall $\to$ Wall: the wall continues along either of the two
    forward edges --- purely geometric, no color choice ($2$ ways).
  \item Wall $\to$ Junction: both forward edges simultaneously become
    wall segments, forming a Y-junction; again purely geometric
    ($2$ ways).
  \item Junction $\to$ Bulk: a junction cannot collapse directly to
    bulk without first resolving to a wall ($0$ ways).
  \item Junction $\to$ Wall: the junction resolves to a simple wall;
    the new region filling the junction must differ from both colors
    on either side of the surviving wall, giving $q-2$ color choices
    ($q-2$ ways).
  \item Junction $\to$ Junction: the Y-junction propagates as a unit
    along one geometric forward direction ($1$ way).
\end{itemize}
Collecting these weights into a transfer matrix:
\begin{equation}
    M_\Delta = \begin{pmatrix} 1 & 1 & 0 \\ q & 2 & 2 \\ 0 & q-2 & 1
    \end{pmatrix},
    \label{eq:Mtri}
\end{equation}
where rows index the destination (next) state and columns index the
source (current) state, consistent with Eq.~\eqref{eq:Msq}.  The
entry $W\!\to\!J=2$ is purely geometric: it counts the two
new wall segments created when both forward edges become active,
independent of $q$.  The entry $J\!\to\!W=q-2$ is where topology and
color multiplicity first couple: resolving a junction into a simple wall
requires choosing a color for the new region that is incompatible with
both colors already present, and there are precisely $q-2$ such choices.
This asymmetry --- entering a junction is geometric, leaving one is
colored --- is the microscopic origin of the algebraic inseparability
discussed in Sec.~\ref{sec:junction}.
The dominant eigenvalue is
\begin{equation}
    \lambda_\Delta = \frac{3+\sqrt{12q-15}}{2},
    \qquad
    s_{\rm step}^{\Delta} = \ln\lambda_\Delta.
    \label{eq:lamtri}
\end{equation}
At $q=2$: $\lambda_\Delta=3$ and with $E_{\rm step}=4J$:
\begin{equation}
    T_c^\Delta = \frac{4J}{\ln3} \approx 3.6410J,
\end{equation}
which coincides with the exact triangular Ising critical
temperature~\cite{Houtappel1950,Wannier1950}.  The corresponding
triangular critical fugacity is
\begin{equation}
    v_T = e^{2J/T_c^\Delta}-1 = e^{(\ln 3)/2}-1 = \sqrt{3}-1.
    \label{eq:vT}
\end{equation}
One can verify that $v_T$ satisfies the exact triangular critical
condition~\cite{Baxter1982}
\begin{equation}
    v_T^3 + 3v_T^2 = q,
    \label{eq:tri_cubic}
\end{equation}
confirming consistency of the matrix result with the known exact
condition at $q=2$.
At $q=3$: $\lambda_\Delta\approx3.791$, giving $T_c\approx3.001J$, a
$5.3\%$ deviation from the known value
$T_c\approx3.170J$~\cite{Ghaemi2004}.  This deviation reflects the
breakdown of the minimal Markovian approximation when Junction--Junction
correlations become extensive.
The characteristic polynomial of $M_\Delta$ is irreducible: it cannot
be written as $(\lambda-m)^2=q^p$ for any $m,p$ independent of $q$.
This algebraic inseparability is the signature of an effective
frustration in the dual-lattice wall variables, arising from the
coupling between lattice topology and color entropy that bipartiteness
eliminates.
\section{Honeycomb Lattice: Consistent Axiom Application}
\label{sec:hc}
The honeycomb ($z=3$) is bipartite, so no Junction state is needed.
It is also the most instructive case in the paper: it is the only common
lattice that is bipartite but \emph{not} self-dual, and applying the
framework axioms consistently to it reveals exactly what self-duality
does for the square lattice.
\subsection{The wrong object: naive counting on the honeycomb}
A naive two-state counting performed directly on the honeycomb geometry,
where a trivalent vertex provides two forward continuations, gives
\begin{equation}
    M_{\rm hc}^{\rm naive} = \begin{pmatrix} 1 & 1 \\ q & 2 \end{pmatrix},
    \qquad
    \lambda_{\rm hc} = \frac{3+\sqrt{1+4q}}{2},
    \label{eq:Mhc}
\end{equation}
with $E_{\rm step}=(z-2)J=J$.  At $q=2$: $\lambda_{\rm hc}=3$,
giving $T_c=J/\ln3\approx0.910J$.  This is not the exact critical
temperature of either the honeycomb or the triangular Ising model.
The counting encoded in $M_{\rm hc}^{\rm naive}$ is geometrically
correct for paths \emph{on the honeycomb}: at each trivalent vertex one
edge arrives and two leave, so $W\!\to\!W=2$ is the right weight for
that lattice.  The error is not in the counting but in the
identification of which geometric object is being counted.  In two
dimensions, a domain wall is a boundary between spin regions; such
boundaries are segments of dual-lattice edges by construction, so
domain-wall paths always live on the \emph{dual} lattice, not on the
spin lattice itself.  The dual of the honeycomb is the triangular
lattice ($z=6$), and it is on the triangular geometry --- with
$E_{\rm step}=(z-2)J=4J$ --- that the counting must be performed.
$M_{\rm hc}^{\rm naive}$ counts the right kind of object (open paths)
on the wrong lattice (honeycomb instead of triangular) with the wrong
energy scale ($J$ instead of $4J$); the result is therefore without
physical meaning.  This geometric mismatch is independent of
Kramers--Wannier duality: the issue is not a transformation of the
partition function but the elementary fact that domain-wall paths
inhabit the dual lattice.  Table~\ref{tab:hc_compare} shows the three
values side by side.
\begin{table*}[t]
\caption{Honeycomb Ising ($q=2$): naive counting, correct triangular
matrix, and exact honeycomb result.  The naive matrix uses the wrong
geometry and energy scale.  The correct two-step procedure uses the
triangular matrix ($E_{\rm step}=4J$) followed by the exact duality
relation $v_H\cdot v_T=q$.}
\label{tab:hc_compare}
\begin{ruledtabular}
\begin{tabular}{ccc}
Method & $\lambda$ or condition & $T_c/J$ \\
\hline
Naive $2\times2$ (wrong lattice, $E_{\rm step}=J$) & $\lambda=3$ & 0.910 \\
Triangular matrix (correct, $E_{\rm step}=4J$)     & $\lambda_\Delta=3$ & 3.6410 \\
Honeycomb (after duality $v_H v_T=q$)              & $e^{2J/T_c}=2+\sqrt{3}$ & 1.5187 \\
\end{tabular}
\end{ruledtabular}
\end{table*}
\subsection{The correct procedure: triangular matrix plus duality}
The axiom-consistent treatment of the honeycomb proceeds in two steps.
\emph{Step 1.}  Domain walls on the honeycomb are paths on the
triangular dual lattice.  The correct transfer matrix is therefore
$M_\Delta$ (Eq.~\eqref{eq:Mtri}), built on the triangular geometry
with $E_{\rm step}=(z-2)J=4J$.  At $q=2$ this gives
$\lambda_\Delta=3$ and
\begin{equation}
    T_c^\Delta = \frac{4J}{\ln 3} \approx 3.6410J,
\end{equation}
which coincides with the exact triangular Ising critical temperature;
the corresponding triangular critical fugacity is
\begin{equation}
    v_T = e^{2J/T_c^\Delta}-1 = \sqrt{3}-1.
\end{equation}
This is not an additional approximation beyond the Markovian framework:
when applied to the correct (triangular) geometry, the construction
reproduces the exact triangular result at $q=2$.
\emph{Step 2.}  The bond fugacity $v = e^{2J/T}-1$
(Eq.~\eqref{eq:fugacity}) is the natural variable of the
Kramers--Wannier duality.  The honeycomb and triangular spin models
are an exact dual pair; their fugacities satisfy~\cite{Baxter1982}
\begin{equation}
    v_H \cdot v_T = q,
    \label{eq:duality}
\end{equation}
exactly.  For $q=2$ this follows directly from the Kramers--Wannier
transformation $e^{-2K^*}=\tanh K$, which gives $v^*=2/v$ and hence
$v\cdot v^*=2=q$.  Using $v_T=\sqrt{3}-1$ from Step~1:
\begin{equation}
    v_H = \frac{q}{v_T} = \frac{2}{\sqrt{3}-1} = \sqrt{3}+1,
    \qquad
    T_c^{\rm hc} = \frac{2J}{\ln(2+\sqrt{3})} \approx 1.5187J,
    \label{eq:Tchc}
\end{equation}
the known exact honeycomb Ising result~\cite{Baxter1982,Fisher1967}.
This two-step procedure is not a patch applied to a failed counting.
It is the axiom-compliant treatment of any non-self-dual lattice: the
transfer matrix is built on the dual geometry, and the duality relation
translates the result back to the spin model.  For the square lattice
these two steps collapse into one because the dual geometry is the same
as the spin-model geometry.  The honeycomb makes the distinction
explicit.
\subsection{Why the ansatz $\lambda=m+q^p$ cannot apply to the honeycomb}
The exact honeycomb critical condition from the star-triangle relation
is~\cite{Baxter1982}:
\begin{equation}
    v_H^3 - 3q\,v_H - q^2 = 0.
    \label{eq:hc_cubic}
\end{equation}
This is the honeycomb equation; the triangular equation is
$v_T^3+3v_T^2=q$ (Eq.~\eqref{eq:tri_cubic}), and the two are related
by $v_H v_T=q$.  Substituting $v=q^p$ into Eq.~\eqref{eq:hc_cubic}
yields
\begin{equation}
    q^{3p} - 3q^{1+p} - q^2 = 0,
    \label{eq:hc_residual}
\end{equation}
requiring $3p=1+p=2$ simultaneously --- forcing $p=2/3$ and $p=1/2$
at once, which is impossible.  No choice of $m$ or $p$ can satisfy
Eq.~\eqref{eq:hc_cubic} globally: the exact honeycomb condition is an
irreducible cubic, while $s_{\rm step}^{\rm ansatz}=\ln(m+q^p)$ is a
one-parameter monomial family belonging to a different function class
entirely.  This is a direct algebraic consequence of the absence of
self-duality, which for the square lattice collapses the critical
condition to the pure monomial $v^2=q$ and makes the ansatz exact.
\subsection{Why the triangular--honeycomb dual pair is not locally symmetric}
The triangular and honeycomb lattices are exact duals, but the
domain-wall counting does not run equally simply in both directions.
For the triangular spin model, walls live on the honeycomb dual, which
is bipartite; this suppresses local color-incompatibility constraints,
but does not eliminate the need for a Junction state when multiple
domains meet.  For the honeycomb spin model, however, walls live on the
triangular dual, which is non-bipartite.  The spin model itself remains
ferromagnetic and unfrustrated, but the wall variables on the dual
triangular lattice acquire an effective frustration: color-compatibility
constraints around local circuits cannot be resolved independently.
This is precisely why a Junction state becomes unavoidable in the
triangular-wall description.  The exact duality between the two spin
models remains true, but the simplicity of a local transfer-matrix
counting does not.
\section{Simple Cubic Lattice: Geometric Extrapolation}
The simple cubic lattice ($z=6$) is bipartite and orthogonal, with no
known exact dual partner in 3D.  The ansatz~\eqref{eq:ansatz} with $m=1$
and $p=1/2$ gives
\begin{equation}
    s_{\rm step}^{\rm SC} \approx \ln(1+\sqrt{q}),
    \qquad
    T_c^{\rm SC} \approx \frac{4J}{\ln(1+\sqrt{q})}.
    \label{eq:Tc_sc}
\end{equation}
The assignment $p=1/2$ is motivated by a geometric argument: in 3D the
domain wall is a 2D surface, and for an orthogonal bipartite lattice the
two independent surface directions contribute independently, giving $m=1$
and $p=1/2$ by the same structural argument as for the square lattice
(Appendix~\ref{app:sc}).
For $q=2$: $s_{\rm step}^{\rm SC}\approx\ln(1+\sqrt{2})\approx0.881$
and $T_c\approx4.538J$, within $0.58\%$ of the numerical value
$4.5115J$~\cite{Ferrenberg1991}.  Table~\ref{tab:sc} shows that the
deviation grows with $q$, reflecting increasing inter-wall correlations.
\begin{table}[t]
\caption{Simple cubic lattice ($z=6$): ansatz versus numerical results.
Numerical values from Refs.~\cite{Ferrenberg1991,Gavai1989}.}
\label{tab:sc}
\begin{ruledtabular}
\begin{tabular}{cccc}
$q$ & $s_{\rm step}=\ln(1+\sqrt{q})$ & $T_c/J$ (ansatz) & $T_c/J$ (numerical) \\
\hline
2 & 0.8814 & 4.5383 & 4.5115 \\
3 & 1.0051 & 3.9794 & $\approx3.78$ \\
4 & 1.0986 & 3.6410 & $\approx3.30$ \\
\end{tabular}
\end{ruledtabular}
\end{table}
\section{Junction State, Frustration, and AFM Entropy}
\label{sec:junction}
The Junction state required on the triangular lattice unifies two
apparently distinct phenomena.
In the FM picture, Junction is a minority excitation above the Bulk
reference; its presence forces the irreducible characteristic polynomial
and limits the accuracy of $s_{\rm step}^{\rm ansatz}$ at $q\geq3$.
In the AFM triangular Ising model, Junction condensation is complete.
Wannier~\cite{Wannier1950} computed the resulting extensive ground-state
entropy exactly, $S_0/N=0.3231\,k_B$.  For the $q=3$ Potts AFM,
Baxter~\cite{Baxter1970} showed the ground state is a dense Junction
tiling with power-law correlations.
The chromatic number $\chi$ provides the unifying language, and also
determines whether $m$ in the ansatz $\lambda=m+q^p$ is a pure
topological integer or an effectively $q$-dependent quantity.
On bipartite lattices ($\chi=2$), within the minimal framework, no
transition weight mixes topology and color multiplicity: geometric
weights such as $W\!\to\!W$ are integers independent of $q$, and $q$
appears only in the color-opening weight $B\!\to\!W$.  As a result $m$
is a pure topological integer, geometry and color entropy are linearly
separable in the growth rate, and the characteristic polynomial factors
as $(\lambda-m)^2=q^p$.
On non-bipartite lattices ($\chi\geq3$) the Junction state introduces
the weight $J\!\to\!W=q-2$, which is simultaneously topological (a
junction-resolution event) and color-dependent (counting colors
incompatible with both domains at the surviving wall).  This mixed
weight makes $m$ effectively $q$-dependent within the minimal
description: one cannot write $\lambda_\Delta$ in the form $m+q^p$ for
any $m$ independent of $q$.  The Junction state \emph{hybridizes}
geometry and color entropy at the level of individual matrix entries,
and the polynomial becomes irreducible as a direct algebraic consequence.
Bipartiteness acts as the topological switch: it excludes the mixed
weight $q-2$, leads to a $q$-independent $m$ within this framework, and
with it the factorable polynomial and absence of extensive AFM
frustration follow simultaneously.
\section{Summary}
Table~\ref{tab:Tc} collects critical temperatures.  Four regimes emerge:
\begin{enumerate}
\item \textbf{Self-dual} (square lattice): $s_{\rm step}=\ln(1+\sqrt{q})$
recovers the exact critical temperature; the ansatz coincides with the
exact monomial critical condition $v^2=q$.
\item \textbf{Triangular lattice} (non-self-dual, non-bipartite):
the Junction matrix gives the exact result at $q=2$; the Markovian
approximation breaks down at $q\geq3$ when Junction--Junction
correlations become extensive.
\item \textbf{Honeycomb lattice} (non-self-dual, bipartite): the
axiom-consistent treatment uses the triangular matrix with
$E_{\rm step}=4J$, giving the exact triangular result at $q=2$.
The honeycomb critical temperature then follows from
$v_H\cdot v_T=q$ --- exact, and the correct second step for any
non-self-dual lattice.
\item \textbf{3D, no exact partner} (simple cubic): the ansatz is a
parameter-free approximation, accurate to $0.6\%$ at $q=2$ and
deteriorating with $q$.
\end{enumerate}
\begin{table}[h]
\caption{Critical temperatures ($J=1$) across all lattices.  Square
results exact for $q\leq4$ by self-duality.  Triangular exact at $q=2$
from the Junction matrix.  Honeycomb exact at $q=2$ via duality, not
directly from the counting matrix (see Sec.~\ref{sec:hc}).
Simple-cubic values are geometric extrapolations.}
\label{tab:Tc}
\begin{ruledtabular}
\begin{tabular}{lcccc}
Lattice & $q=1$ & $q=2$ & $q=3$ & $q=4$ \\
\hline
Square (exact)           & 2.885 & 2.269 & 1.990 & 1.820 \\
Triangular (matrix)      & ---   & 3.6410 & 3.001 & --- \\
Tri.\ known~\cite{Ghaemi2004} & --- & 3.6410 & 3.170 & --- \\
Honeycomb (via duality)  & ---   & 1.5187 & ---   & --- \\
Simple cubic (ansatz)    & ---   & 4.5383 & 3.9794 & 3.6410 \\
SC numerical             & ---   & 4.5115 & $\approx3.78$ & $\approx3.30$ \\
\end{tabular}
\end{ruledtabular}
\end{table}
\section{Conclusion}
\label{sec:conclusion}
We have developed a geometric domain-wall framework for Potts
criticality based on the entropy density per interface step,
$s_{\rm step}=\ln\lambda$, and the coarse-grained free-energy balance
$f_{\rm step}=E_{\rm step}-T s_{\rm step}=0$.  Within this reduced
description, the transition is located by the point at which the
energetic cost of extending an interface is compensated by the
exponential growth rate of allowed interface histories.
The main result is not a replacement for the exact partition-function
methods of Kramers--Wannier, Onsager, or Baxter, but a geometric
reorganization of the problem that makes explicit which lattice
properties control the success or failure of a minimal local counting.
Two such properties emerge independently.  The first is
\emph{self-duality}: on the square lattice, where the dual geometry
coincides with the spin-model geometry, the two-state transfer matrix
reproduces the exact monomial critical condition and hence the exact
critical temperature.  The agreement is not accidental: the counting
produces precisely the algebraic structure required by self-duality.
The second is \emph{bipartiteness}: on bipartite lattices, topology and
color multiplicity remain separable at the level of the minimal transfer
matrix, whereas on non-bipartite lattices a Junction state becomes
unavoidable and couples them irreducibly.
For non-self-dual lattices, the framework must be applied in two steps.
The transfer matrix is first constructed on the dual-lattice geometry
where domain walls propagate, yielding the corresponding dual-lattice
critical fugacity.  The critical point of the spin model is then
obtained from the exact duality relation $v v^*=q$.  In this way, the
triangular--honeycomb pair makes explicit a distinction that is hidden
on the square lattice.  The same analysis also clarifies why the two
directions of the dual pair are not equally simple: the non-bipartite
triangular dual introduces effective frustration in the wall variables
and makes a Junction state unavoidable.
The Junction state is the central topological object of the
non-bipartite problem.  In the ferromagnetic setting it marks the onset
of the irreducible coupling between geometry and color entropy and
limits the accuracy of the minimal counting.  In the antiferromagnetic
setting it connects naturally to the extensive Wannier--Baxter
frustration entropy.  The same topological mechanism thus underlies
both the breakdown of separability in the reduced transfer matrix and
the appearance of extensive frustration entropy in the antiferromagnetic
problem.
The simple-cubic application suggests that the geometric ansatz
$\lambda=m+q^p$ may remain useful beyond two dimensions as a
parameter-free approximation when self-duality is unavailable but
bipartite orthogonal geometry still suppresses the Junction sector.
The ansatz is not a fit: it is a structural consequence of the
separability between geometric propagation and color multiplicity that
holds on bipartite orthogonal lattices.  Its good performance at $q=2$
should be interpreted not as an exact result but as evidence that the
leading interface entropy can in some cases be captured by a
remarkably small state space.
The broader message is modest but, we believe, useful: criticality can
be organized geometrically in terms of the leading entropy of
domain-wall propagation, and the success of this reduced description is
controlled by self-duality, bipartiteness, and the presence or absence
of Junction states.  Exact solutions remain the gold standard.  What
the present framework provides is a map of the minimal geometric
structures that those exact solutions implicitly contain.
\begin{acknowledgments}
Ames National Laboratory is operated for the U.S.\ Department of Energy
by Iowa State University under Contract No.\ DE-AC02-07CH11358.
\end{acknowledgments}
\appendix
\section{Square-lattice counting and the minimal $2\times2$ matrix}
\label{app:square}
The recursion
\begin{align}
    a_n &= a_{n-1} + b_{n-1}, &
    b_n &= q\,a_{n-1} + b_{n-1}
\end{align}
gives integer counts $\Omega_n=a_n+b_n$ whose ratio converges to
$\lambda=1+\sqrt{q}$ (Table~\ref{tab:counts}).  The irrationality of
$\lambda$ is a spectral property of the integer matrix; it is not a
consequence of any single irrational local weight.
\begin{table}[h]
\caption{Integer counts for the square-lattice recursion at $q=2$.
The entropy density $s_{\rm step}=\ln(1+\sqrt{2})\approx0.881$ emerges
only asymptotically.}
\label{tab:counts}
\begin{ruledtabular}
\begin{tabular}{ccccc}
$n$ & $a_n$ & $b_n$ & $\Omega_n$ & $\Omega_n/\Omega_{n-1}$ \\
\hline
0 & 1  & 0  & 1  & --- \\
1 & 1  & 2  & 3  & 3.000 \\
2 & 3  & 4  & 7  & 2.333 \\
3 & 7  & 10 & 17 & 2.429 \\
4 & 17 & 24 & 41 & 2.412 \\
5 & 41 & 58 & 99 & 2.415 \\
$\infty$ & & & & $1+\sqrt{2}\approx2.414$ \\
\end{tabular}
\end{ruledtabular}
\end{table}
\section{Orthogonal stacking and the simple-cubic extrapolation}
\label{app:sc}
In $d=3$ the domain wall is a 2D surface.  For an orthogonal bipartite
lattice the two independent surface directions contribute independently:
each direction contributes 2 opposing domains, giving
$N_{\rm dom}=2^{d-1}=4$ and exponent $p=(d-1)/N_{\rm dom}=1/2$.  The
persistence term remains $m=1$ because the local domain-adjacency graph
around a dual-lattice edge is a cycle $C_4$: the four domains that meet
at a shared edge in the simple-cubic tiling form a ring, not a complete
graph.  Since $C_4$ is bipartite, this local structure preserves the
two-sublattice separation underlying the square-lattice factorization
$(\lambda-1)^2=q$, and the Junction state --- which requires a
non-bipartite local domain graph --- is topologically suppressed.
It is worth noting that Wegner~\cite{Wegner1971} established a duality
between the 3D Ising model on the simple cubic lattice and a
$\mathbb{Z}_2$ lattice gauge theory on the same lattice.  This is
analogous in structure to the Kramers--Wannier duality of the square
lattice, but with a crucial difference: the simple cubic is not
self-dual.  The Ising model maps to a \emph{different} object (a gauge
theory), not to itself, so no self-dual point fixes $T_c$ algebraically.
The 3D critical temperature is not determined by Wegner's duality ---
it requires numerical simulation.  The good performance of the $m=1$,
$p=1/2$ ansatz therefore rests entirely on the geometric
bipartite-orthogonal argument of this appendix, not on any exact duality
constraint.
\section{Exact dual-lattice bond-fugacity relation}
\label{app:duality}
For any planar spin lattice $\mathcal{L}$ with dual $\mathcal{L}^*$,
domain walls are paths on $\mathcal{L}^*$.  The bond fugacity
$v=e^{2J/T}-1$ transforms under this duality as~\cite{Baxter1982}
\begin{equation}
    v \cdot v^* = q,
\end{equation}
an exact result that holds for all $q$ and any dual lattice pair.
\section{An Honest Reckoning}
\label{app:limitations}
This appendix is written in a different register from the rest of the
paper.  It is an attempt to say plainly what this work is, what it
is not, and why the distance between the two matters.
The original motivation behind this project was more ambitious than
what appears in the preceding sections.  The hope was to recast the
criticality of the 2D Potts model as a combinatorial problem: to count
interface microstates exactly, as one counts the states of a gas, and
to derive the critical temperature from first principles without
invoking duality, without transforming the partition function, and
without the machinery of exact integrability.  A two-line combinatorial
proof, transparent and elementary, where Onsager needed pages of
algebra.  That goal was not achieved.  The present paper is what
remained after accepting that it could not be.
\subsection*{What Onsager, Yang, and Baxter actually did}
Lars Onsager's 1944 solution of the square-lattice Ising
model~\cite{Onsager1944} is one of the most remarkable calculations
in the history of theoretical physics.  He found, buried inside the
transfer matrix of a system of interacting spins, the structure of an
infinite-dimensional Lie algebra.  From that algebraic structure the
exact free energy followed, and with it the logarithmic divergence of
the specific heat that had defeated everyone before him.  C.~N.~Yang
subsequently gave a cleaner derivation~\cite{Yang1952}, and Rodney
Baxter extended the reach of these ideas across a vast landscape of
exactly solvable models through the star-triangle (Yang--Baxter)
relation~\cite{Baxter1982}.  What these methods share is that they
work with the full partition function --- the sum over all spin
configurations with correct Boltzmann weights --- and exploit deep
algebraic identities that are properties of that sum, not of any
truncated description of it.  These are not clever tricks;
they are the discovery of genuine mathematical structure in physical
systems, structure that was not put there by the people who found it.
\subsection*{Why the counting approach cannot reach what they reached}
The transfer matrices in this paper act on two or three states.
The transfer matrix in Onsager's solution acts on $2^L$ states for
a row of $L$ spins; its spectrum encodes not just the growth rate of
domain walls but the entire free energy of the system.  The reduction
from $2^L$ to 2 is not a simplification that preserves the essential
physics and discards the rest; it discards precisely the long-range
correlations and loop-closure topology that make the exact solution
possible and that distinguish a phase transition from a crossover.
A domain wall in a finite system is a closed loop, not an open path.
The partition function sums over all closed-loop configurations;
that sum, weighted correctly, generates the algebraic structure
Onsager found.  The minimal matrix counts open paths and treats
closure as an afterthought.  Getting exact results from the closed-loop
ensemble requires the Fortuin--Kasteleyn representation~\cite{Baxter1978}
or the loop-model formulation --- both of which reconstruct, in
different languages, essentially the full partition function.  There
is no shortcut that extracts exact critical exponents or exact
critical temperatures from a two-state local description, except
on the square lattice, where self-duality provides an independent
algebraic pin on $T_c$ that has nothing to do with the counting.
The square-lattice result in this paper is exact for that reason ---
because Kramers and Wannier~\cite{KW1941} gave us the self-dual point
before any counting was done, and the counting merely confirms it.
Attributing the exactness to the transfer matrix would be like
attributing the accuracy of a stopped clock to the quality of its
escapement.
\subsection*{What this paper does instead, and why that is enough}
Having said all of that, these observations are not without value.
The connections that bipartiteness and self-duality are independent
properties, that the Junction state is the common thread connecting FM
corrections and AFM frustration entropy, that the counting must be
performed on the dual-lattice geometry or it gives a wrong answer for
the wrong reason --- these are true, they are not trivial, and they are
not visible from inside the exact solution.  Onsager's calculation tells
you the free energy of the square-lattice Ising model to any desired
precision; it does not tell you why the triangular lattice needs a
Junction state or why the honeycomb and triangular are each other's dual
in the sense relevant to domain-wall entropy.
The exact solvers built cathedrals.  This paper draws a map of the
ground on which they stand.  A map is not a cathedral.  But it is
not nothing either, and maps sometimes show things that cathedrals
obscure.
The author began this project hoping to find a combinatorial
proof that would have impressed Onsager.  What emerged instead is a
geometric picture that might at least have interested him.
That is a smaller thing, but it is an honest one.
\section{Recursive construction of domain-wall configurations}
\label{app:recursive}
To make explicit the origin of the transfer matrix in Eq.~\eqref{eq:Msq},
we formulate the counting of domain-wall configurations as a recursive
process.  The key idea is to construct interface configurations step by
step, while keeping track only of whether a domain wall is currently
active at the endpoint of the construction.
\subsection{State definitions}
We consider sequences of length $n$ representing partial configurations
of a domain wall on the dual lattice.  At each step, the configuration
is characterized by one of two states:
\begin{itemize}
\item \textbf{Bulk state} ($a_n$): no domain wall is currently active
  at the endpoint.  The system is locally in a uniform Potts domain.
\item \textbf{Wall state} ($b_n$): a domain wall is active at the
  endpoint, separating two distinct Potts domains.
\end{itemize}
Thus, $a_n$ and $b_n$ count the number of partial configurations of
length $n$ that terminate in the bulk and wall states, respectively.
\subsection{Allowed local transitions}
The recursion follows from enumerating all allowed local transitions
between these states when extending a configuration by one step:
\begin{enumerate}
\item \textbf{Bulk $\to$ Bulk}: the system remains in a uniform domain
  (1 way).
\item \textbf{Bulk $\to$ Wall}: a domain wall is initiated, with
  multiplicity $q$ associated with selecting a distinct Potts color
  for the new domain ($q$ ways).
\item \textbf{Wall $\to$ Wall}: the domain wall continues without
  closing (1 way).
\item \textbf{Wall $\to$ Bulk}: the domain wall closes and the system
  returns to a uniform region (1 way).
\end{enumerate}
These rules encode only the combinatorics of interface formation and
do not involve energetic or Boltzmann weights.
\subsection{Recursive relations}
\begin{figure*}[t]
\centering
\includegraphics[width=\textwidth]{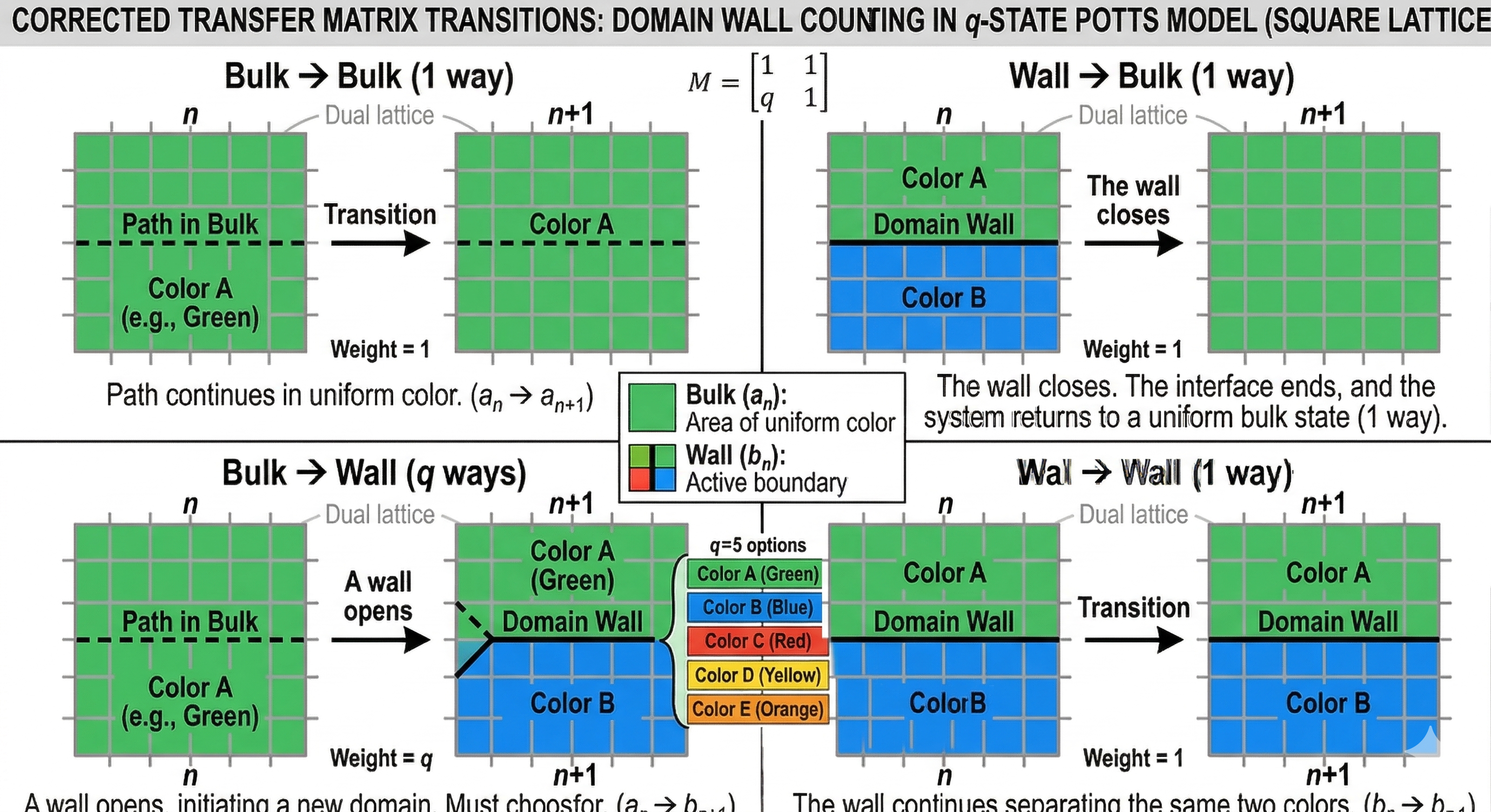}
\caption{All four local transitions of the two-state transfer matrix
$M$ for domain-wall counting on the square lattice dual, illustrated
for $q=5$ Potts colors.  Each panel shows a step of the probe path
(dashed line) on the dual lattice from column $n$ to column $n+1$.
\emph{Bulk} ($B$, green) denotes a path step lying entirely within a
single uniform Potts domain; \emph{Wall} ($W$, boundary between green
and blue) denotes a path step crossing an active domain boundary
separating two distinct Potts colors.
(Top left)~$B\!\to\!B$, weight $1$: the path continues in a uniform
domain; no new color degree of freedom is introduced.
(Top right)~$W\!\to\!B$, weight $1$: the wall closes and the color is
absorbed back into the surrounding uniform domain; only one outcome is
possible.
(Bottom left)~$B\!\to\!W$, weight $q$: a new domain wall opens and the
newly created domain selects one of the $q$ available Potts colors (five
options shown); this is the sole source of color multiplicity in the
counting.
(Bottom right)~$W\!\to\!W$, weight $1$: the wall continues separating
the same two colors; no new color choice is made.
These weights define the recursion $a_{n+1}=a_n+b_n$,
$b_{n+1}=q\,a_n+b_n$, and the transfer matrix
$M=\bigl(\begin{smallmatrix}1&1\\q&1\end{smallmatrix}\bigr)$,
whose dominant eigenvalue $\lambda=1+\sqrt{q}$ gives the entropy per
step $s_{\rm step}=\ln(1+\sqrt{q})$.
}
\label{fig:BW_schematic}
\end{figure*}

From the above transitions, the number of configurations at step $n+1$
follows:
\begin{align}
a_{n+1} &= a_n + b_n, \\
b_{n+1} &= q\,a_n + b_n.
\end{align}
The first equation reflects that a bulk endpoint arises either by
remaining in bulk or by closing a wall.  The second reflects that a
wall endpoint arises either by opening a new wall from the bulk (with
multiplicity $q$) or by continuing an existing wall.
These relations can be written in matrix form:
\begin{equation}
\begin{pmatrix}
a_{n+1} \\
b_{n+1}
\end{pmatrix}
=
\begin{pmatrix}
1 & 1 \\
q & 1
\end{pmatrix}
\begin{pmatrix}
a_n \\
b_n
\end{pmatrix}.
\end{equation}
See Fig.\ \ref{fig:BW_schematic} for illustration of creating the matrix. 
\subsection{Initial condition and asymptotic growth}
Starting from a uniform configuration with no active wall,
\begin{equation}
a_0 = 1, \qquad b_0 = 0,
\end{equation}
the total number of configurations after $n$ steps is
\begin{equation}
N_n = a_n + b_n.
\end{equation}
For large $n$, the growth is exponential,
\begin{equation}
N_n \sim \lambda^n,
\end{equation}
where $\lambda$ is the largest eigenvalue of the transfer matrix:
\begin{equation}
\lambda = 1 + \sqrt{q}.
\end{equation}
This defines an entropy per step
\begin{equation}
s_{\mathrm{step}} = \ln(1+\sqrt{q}),
\end{equation}
which enters directly into the estimate of the critical temperature.
\subsection{Eigenvalues and eigenvectors: asymptotic structure}
The recursive relations are governed by the transfer matrix
\begin{equation}
M =
\begin{pmatrix}
1 & 1 \\
q & 1
\end{pmatrix}.
\end{equation}
The asymptotic behavior of the system is controlled by the eigenvalues
of $M$, obtained from the characteristic equation
\begin{equation}
\det(M - \lambda I) = 0,
\end{equation}
which yields
\begin{equation}
\lambda_\pm = 1 \pm \sqrt{q}.
\end{equation}
The total number of configurations grows exponentially as
\begin{equation}
N_n \sim \lambda_+^n = (1+\sqrt{q})^n,
\end{equation}
giving an entropy per step
\begin{equation}
s_{\mathrm{step}} = \ln(1+\sqrt{q}).
\end{equation}

Beyond determining the growth rate, the eigenvectors of $M$ encode the
internal structure of the domain-wall ensemble. The right eigenvectors
satisfy $M v_\pm = \lambda_\pm v_\pm$ and can be written as
\begin{equation}
v_+ =
\begin{pmatrix}
1 \\
\sqrt{q}
\end{pmatrix},
\qquad
v_- =
\begin{pmatrix}
1 \\
-\sqrt{q}
\end{pmatrix}.
\end{equation}

The dominant eigenvector $v_+$ determines the asymptotic composition of
configurations. In particular, for large $n$,
\begin{equation}
\frac{b_n}{a_n} \;\longrightarrow\; \sqrt{q},
\end{equation}
showing that the leading entropy-carrying configurations have a fixed
ratio of wall to bulk endpoints. Thus, while the microscopic rule for
initiating a wall carries a multiplicity $q$, the recursive dynamics
renormalizes this into an effective weight $\sqrt{q}$ governing the
large-scale structure of the interface ensemble.

The subleading eigenvector $v_-$ describes the fluctuation mode about
this asymptotic composition. Writing the initial condition
\begin{equation}
\begin{pmatrix}
1 \\
0
\end{pmatrix}
= \frac{1}{2} v_+ + \frac{1}{2} v_-,
\end{equation}
the full solution can be expressed as
\begin{equation}
\begin{pmatrix}
a_n \\
b_n
\end{pmatrix}
=
\frac{1}{2} \lambda_+^n v_+ +
\frac{1}{2} \lambda_-^n v_-,
\end{equation}
or explicitly,
\begin{align}
a_n &= \frac{1}{2}\left(\lambda_+^n + \lambda_-^n\right), \\
b_n &= \frac{\sqrt{q}}{2}\left(\lambda_+^n - \lambda_-^n\right).
\end{align}
The second eigenvalue $\lambda_-$ therefore controls the finite-size
corrections and the rate at which the ratio $b_n/a_n$ approaches its
asymptotic value.

In this way, the full spectrum of the transfer matrix provides a
complete description of the interface statistics: the dominant
eigenvalue sets the entropy per step, while the eigenvectors determine
the internal balance between bulk and wall configurations and the
structure of fluctuations about this balance.
\subsection{Interpretation and relation to cluster representations}
It is important to emphasize that this construction does not explicitly
enumerate closed domain-wall loops.  Instead, it counts \emph{partial
interface histories}, with closed loops appearing implicitly as sequences
in which a wall is opened and subsequently closed.
The multiplicity factor $q$ associated with the Bulk $\to$ Wall
transition reflects the freedom to select a distinct Potts color when
an interface is created.  In this sense, the recursion captures, at a
minimal level, the same combinatorial ingredient that appears in cluster
formulations of the Potts model, where each connected domain carries a
weight proportional to $q$.
While the present approach does not rely on an explicit
Fortuin--Kasteleyn construction, the emergence of the factor $q$ in the
recursion is consistent with the interpretation that interface formation
and domain counting are intrinsically linked.  The success of this
minimal description for the square lattice suggests that the dominant
contribution to the entropy is governed by this balance between
interface propagation and domain multiplicity, rather than by detailed
geometric constraints of individual domain-wall loops.

\end{document}